%% file: ms.tex
\documentclass[preprint2]{aastex6}
\usepackage{natbib}
\usepackage{bethmacros}
\begin{document}

\title{$\Lambda CDM$ is Consistent with SPARC Radial Acceleration Relation}

\author{B.W. Keller\altaffilmark{1} and J. W. Wadsley}
\affil{Department of Physics and Astronomy, McMaster University, Hamilton,
Ontario, L8S 4M1, Canada}

\altaffiltext{1}{kellerbw@mcmaster.ca}

\begin{abstract}
    Recent analysis \citep{McGaugh2016} of the SPARC galaxy sample found a
    surprisingly tight relation between the radial acceleration inferred from
    the rotation curves, and the acceleration due to the baryonic components of
    the disc.  It has been suggested that this relation may be evidence for new
    physics, beyond $\Lambda CDM$.  In this letter we show that 32
    galaxies from the MUGS2 match the SPARC acceleration relation. These
    cosmological simulations of star forming, rotationally supported discs were
    simulated with a {\sc WMAP3} $\Lambda CDM$ cosmology, and match the SPARC
    acceleration relation with less scatter than the observational data.  These
    results show that this acceleration relation is a consequence of
    dissipative collapse of baryons, rather than being evidence for exotic
    dark-sector physics or new dynamical laws.
\end{abstract}


\section{Introduction}
For nearly a century, observations of kinematics in galaxies and clusters of
galaxies have found large velocities inconsistent with the luminous matter
within them.  Even when thorough, comprehensive surveys of the baryonic mass
within galaxies and clusters have been performed, most of the matter has been
found to be missing.  \citet{Zwicky1937} presented observations of galaxy
velocity dispersions in the Coma cluster, and proposed that the bulk of that
cluster's mass was some sort of dark matter (DM).  Later, the groundbreaking
observations of \citet{Rubin1970} showed that this dark matter was also
ubiquitous within disc galaxies like our own.  Today, there is a wealth of
evidence for cold dark matter, not just from galaxy kinematics, but from the
formation of large-scale structure \citep{Blumenthal1984}, the cosmic microwave
background power spectrum \citep{Planck2014},  and the primordial abundances of
elements after Big Bang Nucleosynthesis \citep{Walker1991}.  Dark matter is now
part of the standard cosmology, $\Lambda CDM$, in which most of the
matter in our universe is in fact dark.  Despite this, we still do not know the
actual form that dark matter particles take.  Both direct detection experiments
and searches for dark matter annihilation have failed to conclusively observe
these particles \citep{Aprile2012}, and as such, alternative explanations for
the kinematics of galaxies have been proposed.

The Spitzer Photometry \& Accurate Rotation Curves (SPARC) sample, presented in
\citet{Lelli2016b} is a new set of observations and derived mass models for a
large number of rotation-dominated galaxies.  By using $3.6\mu m$ observations,
the stellar mass can be estimated with great accuracy.  The stellar mass is
complemented with $21 cm$ observations of {\sc HI} to get a measure of the gas
mass within the disc. The recent paper by \citet{McGaugh2016} analyzed this
sample, and determined a relation between the observed
radial acceleration determined from the rotation curve ($g_{obs}$), and the
acceleration induced by the baryons observed in the disc ($g_{bar}$).
\citet{McGaugh2016} found that for large values of $g_{bar}$, $g_{obs}\sim
g_{bar}$, while for values of $g_{bar} \lesssim 10^{-10} m\;s^{-2}$, the
observed acceleration begins to rapidly outstrip the acceleration one would
expect from the observed baryons.  They find that the relation between $g_{bar}$
and $g_{obs}$ is well fit by:
\begin{equation} \label{g_fit}
    g_{obs} = \frac{g_{bar}}{1-\exp{(-\sqrt{g_{bar}/g_\dagger})}},
\end{equation}
where $g_\dagger=1.20\pm 0.26\times10^{-10}\;ms^{-2}$.  In addition to the
simple functional form, \citet{McGaugh2016} find a surprisingly low scatter in
this relation, with residuals normally distributed with $\sigma =
0.11\;\mathrm{dex}$.  The authors noted that this is the same functional form as
the Modified Newtonian Dynamics (MOND) \citep{Milgrom1983} acceleration law,
which attempts to explain galaxy rotation curves without DM.

A correlation between the total acceleration  seen in disc galaxies and the
acceleration due only to baryons has been known for some time
\citet{Sancisi2004,McGaugh2004}.  Until recently, this has primarily been
examined through the Mass-Discrepancy Acceleration Relation (MDAR): $g_{bar}$
vs. $M_{tot}/M_{bar}$.  \citet{McGaugh2016} directly probes a more fundamental
relation, the Radial Acceleration Relation (RAR), with number of
improvements that reduce the observational uncertainties.

In discussing these results, \citet{McGaugh2016} offer three possible
explanations for the tight relation. 
\begin{enumerate}
    \item The end point of galaxy formation with conventional (baryonic?)
        physics.
    \item New dark sector physics coupling dark matter and baryons
    \item New dynamical laws (such as MOND
        Tensor-Vector-Scalar Gravity (TeVeS)
        \citep{Bekenstein2004}, etc.)
\end{enumerate}

This is not the first set of observations that appear to be in discordance with
$\Lambda CDM$.  N-body simulations of halo formation have found DM halos follow
a universal, ``cuspy'' density profile \citep{Navarro1996}.  Yet observations of
dwarf galaxies in the local universe find flat, ``cored'' central densities (the
``cusp-core problem'', \citealt{Walker2011}).  Meanwhile, DM-only simulations
were finding that the local group should contain thousands of dwarfs, in
contrast to the dozens actually observed (the ``missing satellites problem''
\citealt{Klypin1999}).  Many of these halos are large enough that suppression of
star formation by reionization could not explain their absence from the
observations (the ``too big to fail'' problem \citealt{Boylan-Kolchin2011}).  

A common feature in each of these conflicts is the comparison of observations to
simulations of galaxy formation that rely purely on N-body, DM-only simulations.
We now know that the impact of baryonic physics, chief among them the feedback
from massive stars and black holes, can have a dramatic effect on the star
formation history (e.g. \citealt{Keller2015}) and density profile of galaxies
\citep{Mashchenko2006}.  Multiple studies \citep[etc]{Pontzen2012,Sawala2016}
have found these problems disappear when galaxies are simulated with gas
dynamics, along with reasonable models for star formation, radiative cooling,
and stellar feedback.  This is what constitutes a modern theory of galaxy
formation, the first of the three options offered to explain the 
RAR.  Galaxies are formed through the gravitational collapse
of collisional particles (gas) into a rotationally supported disc.  Conservation
of angular momentum, combined with star formation and feedback within that disc,
leads to the observed scaling relations and galaxy properties we see today.
Whether this can also reproduce the RAR has been yet to
be demonstrated.

In this letter, we show that the apparent tension between models of galaxy
formation in $\Lambda CDM$ and the SPARC observations also evaporates when the
collisional collapse of baryons is taken into account.  We find that the
$g_{obs}-g_{bar}$ relation for a set of pre-existing cosmological galaxy
simulations, evolved in a conventional $\Lambda CDM$ cosmology, matches the
SPARC acceleration relation, with even tighter scatter than the observed sample.

\section{The MUGS2 Sample}
The McMaster Unbiased Galaxy Simulations 2 (MUGS2) sample is an unbiased,
statistically representative set of 18 cosmological zoom-in simulations of $L*$
disc galaxies.  These galaxies were simulated in a {\sc WMAP3} $\Lambda CDM$
cosmology, with parameters $H_0 = 73\kms \Mpc^{-1}$, $\Omega_M=0.24$,
$\Omega_{bar}=0.04$, $\Omega_\Lambda=0.76$, and $\sigma_8=0.76$.  The MUGS2
$z=0$ halo masses range from $3.7\times10^{11}\Msun$ to $2.2\times10^{12}\Msun$,
with disc masses ranging from $1.8\times10^{10}\Msun$ to
$2.7\times10^{11}\Msun$.  For more details on the creation of the MUGS2 initial
conditions, see the original MUGS paper, \citet{Stinson2010}.  For more
information on the simulations themselves, see \citet{Keller2015,Keller2016}.

MUGS2 was simulated using the modern smoothed particle hydrodynamics code {\sc
Gasoline} \citep{Wadsley2004,Keller2014}.  The simulations used metal line
radiative cooling \citep{Shen2010}, as well as a simple Schmidt law for star
formation.  What sets MUGS2 apart from the original MUGS, aside from improved
hydrodynamics, is the use of a physically motivated, first principles model for
treating feedback from supernovae. (SNe).  Originally presented in
\citet{Keller2014}, the superbubble model captures the effects of thermal
conduction and evaporation between a hot, SNe heated bubble and a surrounding
shell of cold, swept-up interstellar medium (ISM).  This model was derived to allow unresolved
superbubbles to radiatively cool at realistic rates, with no free parameters,
while automatically capturing the effects of clustered SNe.

In addition to the central spirals, we also include a number of dwarf companions
from the MUGS2 sample.  As with \citet{McGaugh2016}, we exclude galaxies that
are experiencing significant tidal interactions.  \citet{Joshi2016} showed that
tidal interactions on infalling galaxies can occasionally be seen out to 3
virial radii, we select galaxies between 3 to 5 virial radii from the central
spiral. We exclude halos beyond 5 virial radii because our zoom-in simulations
do not contain gas particles at these distances. In order to limit the effects
of poor resolution, and to ensure that each radial bin contains sufficient
baryonic resolution, we select only galaxies that contain 100 or more star
particles.  This gives us an additional 14 galaxies at $z=0$, for a total of
32 galaxies.  The stellar, gas, and total virial masses for each of our galaxies
is shown in table~\ref{z0_table}.
\begin{table}
\input{figures/z0_table.tex}
\caption{Redshift 0 properties of our simulated galaxies.  All masses are in
    solar masses. Subscript 0 denotes the central galaxy.}
\label{z0_table}
\end{table}

\subsection{Calculating Accelerations from MUGS2}
In order to compare to the SPARC sample, we located the central halos using the
AMIGA halo finder \citet{Knollmann2009}.  We center the halos using the
shrinking sphere method described in \citet{Power2003}.  Next, in order to
measure rotation curves of the galaxies face-on, we calculate the net angular
momentum vector of all gas within $10\kpc$ of the center of the disk, and rotate
our simulations such that this vector is orthogonal to the x-y plane.
Accelerations were measured in 100 circular annuli $300\pc$ thick.  For dwarfs,
we use 15 $600\pc$-thick annuli, as the dwarfs have much smaller scale lengths,
and to avoid issues from poor sampling of the dark matter or baryon particles
within the dwarfs.  Accelerations were then calculated using a direct N-body
summation on all of the particles in the halo on those particles within the
annulus.  Only the in-plane component of the acceleration was used, to better
follow \citet{McGaugh2016}.  For $g_{obs}$ (the observed acceleration), all
particles (gas, stars, and DM) within the simulation were used.  To calculate
$g_{bar}$, we simply calculate the contributions from stars and gas, $g_*$ and
$g_{gas}$, so that $g_{bar}=g_*+g_{gas}$.  For each of $g_*$ and $g_{gas}$, we
use a direct summation only on those particles (stars and gas respectively).
This process of direct summation to calculate gravity is equivalent to the
numerical solution to Poisson's equation used in \citep{McGaugh2016}.  The mass
model in SPARC \citep{Lelli2016b} included stellar masses estimated from
$3.6\;\mu\rm{m}$ near infrared observations, and gas masses estimated using
$21\;\rm{cm}$ observations of {\sc HI}.  These {\sc HI} masses were converted to
total gas masses using the simple equation $M_{gas}=1.33M_{HI}$.  Rather than
using the total gas mass from our simulations, we follow the {\sc HI}-based
estimate from SPARC by calculating accelerations due to gas using $1.33M_{HI}$,
rather than $M_{gas}$.  This is especially important near the outskirts of the
galaxy, where the contribution to the baryonic mass from ionized gas in the ISM
and circumgalactic medium is most significant.  The HI fraction is calculated
using the radiative cooling code within {\sc Gasoline}, which relies on
tabulated equilibrium cooling rates from {\sc CLOUDY} \citep{Ferland2013}.

\begin{figure}
    \includegraphics[width=0.5\textwidth]{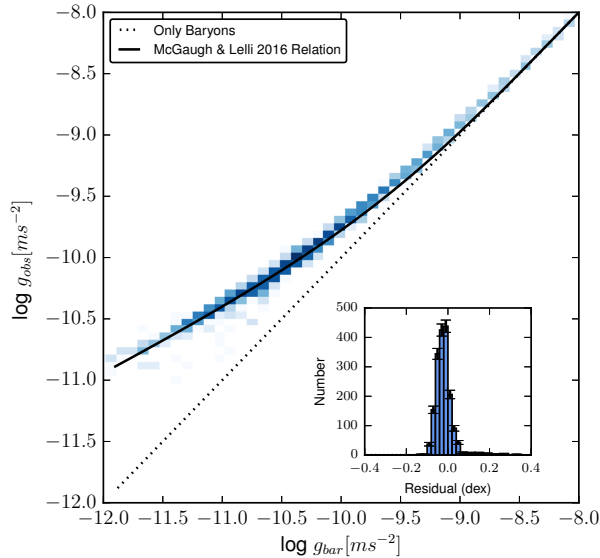}
    \caption{Total acceleration ($g_{obs}$) vs acceleration due to baryons
    ($g_{bar}$) from 2100 data points in the $z=0$ MUGS2 sample, shown in the
    blue 2-dimensional histogram.  The dotted black curve shows the 1:1 relation
    expected if the acceleration was due to baryons alone (without dark matter),
    while the solid line shows the relation presented in \citet{McGaugh2016}.
    A Gaussian distribution fitted to these residuals finds a
    variance of $\sigma=0.06$ dex, significantly lower than the 0.11 dex found
    by \citet{McGaugh2016}.} \label{SPARC_plot}
\end{figure}
\begin{figure}
    \includegraphics[width=0.5\textwidth]{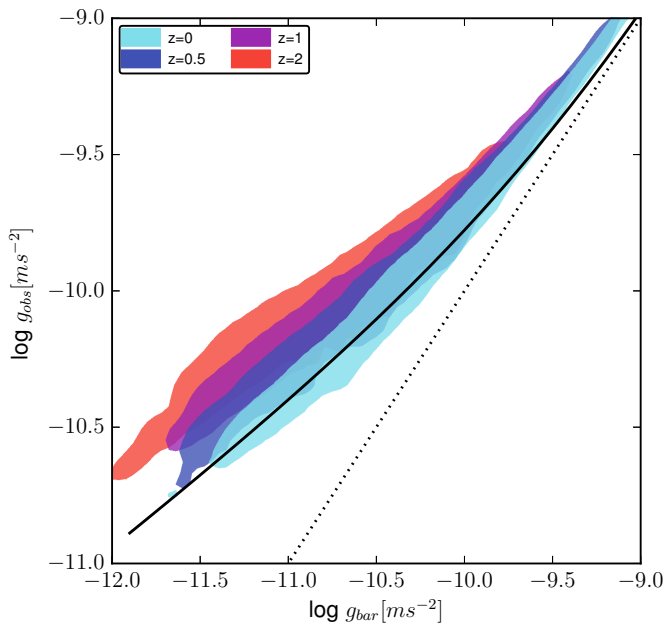}
    \caption{The  simulated $g_{obs}-g_{bar}$ relation is not constant with
    redshift.  As this figure shows, at higher redshift the low $g_{bar}$ slope
    is much shallower than at $z=0$.  This shows that for high redshift
    galaxies, their discs can be depleted of baryons compared with $z=0$.  We
    have focused on the low $g_{bar}$ end of the relation here, where the
    changes are most significant.}
    \label{redshift_evolution}
\end{figure}
\begin{figure}
    \includegraphics[width=0.5\textwidth]{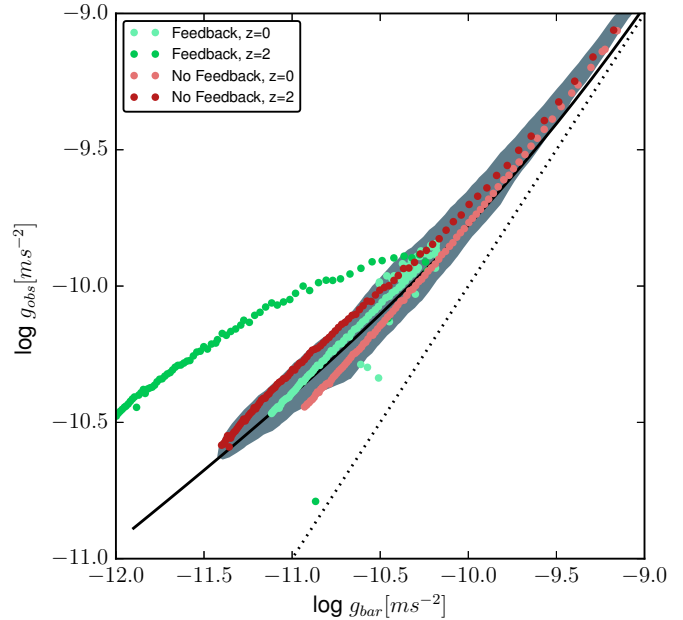}
    \caption{The evolution seen in figure~\ref{redshift_evolution} is primarily
    driven by feedback.  This can be seen when looking at the same galaxy with
    and without feedback.  Without feedback, the baryon fraction within the disc
    increases slightly from $z=0$ to $z=2$, but still roughly follows the 
    RAR.  At $z=2$, strong outflows in the galaxy expel most of the baryons
    from the disc, flattening the acceleration relation.  This effect is
    sensitive to the frequent merger-driven starbursts at high redshift, which
    can drive bursty outflows.}
    \label{FB_effects}
\end{figure}
\section{Results}
\subsection{z=0 Acceleration Relation}
The MUGS2 sample gives us 2100 acceleration data points, just over 3/4 the
sample size of \citet{McGaugh2016}. Figure~\ref{SPARC_plot} shows the
$g_{obs}-g_{bar}$ relation for the MUGS2 sample, compared both to the pure
baryonic acceleration and the RAR.  It is clear these
simulated galaxies follow the \citet{McGaugh2016} relation {\it extremely} well.
As can be seen from the inset residual distribution, our simulated galaxies
follow the SPARC RAR even more tightly than the actual observational data.
The scatter in our results, with $\sigma=0.06$ dex, is consistent with the
\citet{McGaugh2016} estimates.  They decomposed their scatter of $0.11$
dex into different sources, and when all of the observational uncertainties are
removed, the remaining intrinsic scatter gives a variance of $\sigma=0.06$ dex,
very close to the value presented here here.  A reduced $\chi^2$ statistic of the SPARC
relation fit to the $z=0$ MUGS2 data finds a very good fit, with $\chi^2_\nu =
1.25$.  These simulation data are fit by equation~\ref{g_fit} at least as well
as the original SPARC data.

\subsection{Feedback \& the Evolution of the Acceleration Relation}
If the SPARC acceleration relation is in fact due to new physics, it would be
surprising if the relation did not hold at all redshifts.  This would not be the case
if the relation was simply a consequence of galaxy evolution.  In
figure~\ref{redshift_evolution} we show that the acceleration relation in the
MUGS2 sample actually shows significant redshift
dependence, and only settles to the equation~\ref{g_fit} relation near z=0.  For
these data points, we scaled the thickness of the annuli by the cosmic scale
factor $a$, so that $\delta r = 300/(1+z)\pc$. This scaling ensures we are
sampling primarily from the stellar disc, and not well beyond it.  Omitting this
scaling has little effect on these results, save for extending the points to
very low values of $g_{bar}$ and removing points from the high $g_{bar}$ end.
This evolution is a consequence of the huge impact that stellar feedback has on
galaxies at $z\sim2$.  \citet{Keller2015} showed that SNe drive hot outflows
from high redshift galaxies with mass loadings of $\dot M_{out}/\dot M_* \sim
10$.  This leads to discs at high redshift with baryon fractions depleted
relative to those at low redshift.  This feedback effect is clear when a single
galaxy, g1536, is compared to the same galaxy simulated without SNe feedback.
As figure~\ref{FB_effects} shows, the redshift trend is nearly nonexistent
without SNe feedback.  Even at $z=2$, the galaxy without feedback falls within
the scatter of the SPARC observations, and within the scatter of the $z=0$
MUGS2 relation.  This tells us that we need not invoke feedback processes to
explain the $z=0$ SPARC RAR.  Simple dissipational collapse of gas is
sufficient to produce a similar relation.  The evolution as a function of
redshift is therefore dominated primarily by the stronger effect of feedback at
higher redshift.

\section{Discussion}
This paper is the first work to show that the RAR's acceleration scale and tight
scatter, as reported by \citet{McGaugh2016}, can be arise from fully
self-consistent hydrodynamical simulations.

Efforts have been made in the past to explain related acceleration relations
(the MDAR, the Baryonic Tully Fisher (BTFR) \citep{McGaugh2000}, etc.) with
analytic arguments \citep{VanDenBosch2000,Kaplinghat2002}, existing scaling
relations \citep{DiCintio2016}, or hydrodynamical simulations
\citep{SantosSantos2016}.  Analytic studies have found that the MOND-like
scaling relations can arise as a consequence of exponential discs living within
an an NFW halo \citep{VanDenBosch2000}.  \citet{VanDenBosch2000} showed that
this can explain not only Tully-Fisher relation's relatively tight scaling over
a large range of masses and surface densities \citep{McGaugh1998}, but even
explain the appearance of a characteristic acceleration in disk galaxy rotation
curves \citep{McGaugh1999}.

Subsequent studies have suggested that an even harder constraint for $\Lambda
CDM$ to match is the tight scatter in the MDAR \citep{Wu2015} or the RAR
\citep{Milgrom2016}.  A semi-empirical model recently published by
\citet{DiCintio2016} showed that both the BTFR and the MDAR can arise as a
result of galaxies that follow a handful of scaling relations both for the
baryonic content of the galaxy, as well as the dark matter halo it resided
within. Both their model matches to the MDAR and the BTFR did show slightly
higher scatter than the \citet{Lelli2016a} observations (0.17 vs.  0.11 (0.06
intrinsic) dex), which they suggest may be a result of the BTFR's sensitivity to
measuring radius.

The BTFR and MDAR were also examined using galaxies from the MaGICC
\citep{Stinson2013} and CLUES \citep{Gottloeber2010} simulations by
\citet{SantosSantos2016}.    Their results also showed similar scatter to the
MDAR reported in \citet{McGaugh2014}, with a scatter of
$\sigma\sim0.3\;\mathrm{dex}$.\footnote{While this number is not reported in
\citet{McGaugh2014}, we have confirmed this value with Dr. McGaugh in private
communication.}  This is significantly higher than the intrinsic scatter of
$0.06\;\mathrm{dex}$ reported in \citet{McGaugh2016}.  This may be due to the
use of a sample composed of simulations run using different subgrid physics
prescriptions, which will naturally differ from one another.  While matching the
scatter of previous observations, which suffered from much higher observational
uncertainties, matching both the fitting function and small intrinsic scatter of
\citet{McGaugh2016} has never been done prior to this study.

Concurrently with the work presented here, \citet{Ludlow2016} presented a study
using simulations from the EAGLE \citep{Schaye2015} and APOSTLE
\citep{Sawala2016} projects.  EAGLE and APOSTLE use a common set of subgrid
physics.  They found their simulations were fit well by the \citet{McGaugh2016}
functional form of the RAR with $g_\dagger = 3\times10^{-10}\;ms^{-2}$ (well
outside the uncertainties reported in \citet{McGaugh2016} for the value of
$g_\dagger$).  They also found that the z=0 relation is only somewhat sensitive
to the subgrid physics model (as we have found as well).  The fact that their
$g_\dagger$ is larger than ours means EAGLE galaxies are somewhat
baryon-depleted compared to ours.  This, coupled with the lack of redshift
evolution, suggests that the EAGLE feedback model drives stronger outflows at
low redshift compared to the superbubble model used in MUGS2.

The EAGLE subgrid physics model is complex, and involves a number of different
purely numerical parameters that were tuned to reproduce the observed stellar
mass to halo mass relation (SMHMR) and size-mass relation, the details of which
can be found in \citet{Crain2015}.  MUGS2 instead used a well-constrained,
physically motivated model for SN feedback \citep{Keller2014}, with no free
parameters beyond the energy available per supernovae, and which captures the
effects of thermal evaporation that are ignored in EAGLE.  This allows us to
better capture the real variation that occurs in the efficiency of outflows over
cosmic time \citep{Keller2015,Muratov2015}.  Perhaps the clearest conclusion
that can be drawn from the results of this paper and those of \citet{Ludlow2016}
is that the scatter in the RAR will be fairly small regardless of the details of
baryonic process like cooling, star formation, and feedback.  However, the
actual value of $g_\dagger$ {\it is} sensitive to these details, and the RAR may
therefore be a useful new tool for constraining subgrid physics in galaxy
simulations.

\section{Conclusion}
We have shown here that the SPARC RAR can be produced by conventional galaxy
formation in a $\Lambda CDM$ universe.  While we have used a pre-existing set of
simulations, MUGS2, we expect a larger sample designed to match SPARC should
find similar results.  Neither the particular functional form
(equation~\ref{g_fit}) nor the small scatter about this relation requires
anything beyond the dissipational collapse of baryons in a DM halo.  We predict
the fit observed at $z=0$ will not hold at all redshifts: vigorous feedback at
high redshift acts to scour protogalaxies of their baryons, reducing the baryon
fraction of the disc, flattening the $g_{obs}-g_{bar}$ relation.  Stellar
feedback is an essential process if we are to produce realistic galaxies.  In
order for a single RAR to hold at all redshifts, feedback efficiencies would
have to be so low as to produce galaxies with stellar masses and bulge fractions
in conflict with the observed SMHMR, and the observed kinematics of local
galaxies. If one wished to use equation~\ref{g_fit} to fit galaxies at all
epochs, $g_\dagger$ would need to have a significant redshift dependence.  If,
on the other hand, high redshift observations of the $g_{obs}-g_{bar}$ relation
found no evolution in shape, or a steeper slope at low $g_{bar}$, this would in
fact constitute a serious disagreement with $\Lambda CDM$, as it would be
difficult to produce the observed low cosmic star formation efficiency without
strong outflows removing baryons from high redshift discs.

As figure~\ref{FB_effects} shows, the $z=0$ SPARC relation is {\it not} a result of
stellar feedback.  While feedback does change the relationship at high redshift,
its general form is reproduced by simple gas collapse and radiative cooling.
This is one of the few apparent problems in $\Lambda CDM$ that {\it doesn't}
require feedback for its resolution!

\section*{Acknowledgements}
We thank Hugh Couchman, Stacy McGaugh, and Laura Parker for valuable discussion and suggestions.
The simulations used here were performed on \textsc{scinet}, part of
ComputeCanada.  We appreciate these computing allocations.  We also thank NSERC
for funding support.

\bibliography{references}
\end{document}

%% file: figures/z0_table.tex
\begin{tabular}{rrrr}
	\hline
    Galaxy & $M_*$ & $M_{\rm gas}$ & $M_{\rm vir}$ \\
	\hline
	\hline
    $g15807_3$ & $1.66\times10^7$    & $1.52\times10^8$    & $1.74\times10^{10}$ \\
    $g8893_1 $ & $2.14\times10^7$    & $1.06\times10^7$    & $1.10\times10^{10}$ \\
    $g1536_2 $ & $2.62\times10^7$    & $1.03\times10^8$    & $8.36\times10^{9}$ \\
    $g3021_1 $ & $3.24\times10^7$    & $3.93\times10^7$    & $2.49\times10^{10}$ \\
    $g4145_3 $ & $4.04\times10^7$    & $9.55\times10^7$    & $2.83\times10^{10}$ \\
    $g7124_1 $ & $4.76\times10^7$    & $2.83\times10^8$    & $3.54\times10^{10}$ \\
    $g27491_1$ & $6.66\times10^7$    & $4.22\times10^8$    & $2.53\times10^{10}$ \\
    $g1536_1 $ & $1.19\times10^8$    & $3.75\times10^8$    & $5.29\times10^{10}$ \\
    $g4145_2 $ & $1.76\times10^8$    & $8.40\times10^8$    & $4.83\times10^{10}$ \\
    $g15807_2$ & $3.04\times10^8$    & $1.09\times10^9$    & $5.50\times10^{10}$ \\
    $g4145_1 $ & $3.93\times10^8$    & $5.29\times10^8$    & $1.30\times10^{11}$ \\
    $g15807_1$ & $7.54\times10^8$    & $1.68\times10^9$    & $8.17\times10^{10}$ \\
    $g4720_1 $ & $1.03\times10^9$    & $1.36\times10^9$    & $9.70\times10^{10}$ \\
    $g22437_1$ & $1.94\times10^9$    & $8.22\times10^8$    & $1.53\times10^{11}$ \\
    $g7124_0 $ & $5.22\times10^9$    & $4.97\times10^{10}$ & $3.66\times10^{11}$ \\
	$g8893_0 $ & $7.36\times10^9$    & $9.10\times10^{10}$ & $5.80\times10^{11}$ \\
	$g5664_0 $ & $9.44\times10^9$    & $7.29\times10^{10}$ & $4.77\times10^{11}$ \\
	$g21647_0$ & $1.18\times10^{10}$ & $1.01\times10^{11}$ & $7.44\times10^{11}$ \\
	$g422_0  $ & $1.51\times10^{10}$ & $1.24\times10^{11}$ & $7.62\times10^{11}$ \\
	$g28547_0$ & $1.59\times10^{10}$ & $1.67\times10^{11}$ & $9.85\times10^{11}$ \\
	$g1536_0 $ & $1.86\times10^{10}$ & $1.04\times10^{11}$ & $6.49\times10^{11}$ \\
	$g24334_0$ & $2.55\times10^{10}$ & $1.53\times10^{11}$ & $1.02\times10^{12}$ \\
    $g3021_0 $ & $3.63\times10^{10}$ & $1.51\times10^{11}$ & $9.78\times10^{11}$ \\
	$g19195_0$ & $7.15\times10^{10}$ & $9.34\times10^{10}$ & $1.01\times10^{12}$ \\
	$g22437_0$ & $9.03\times10^{10}$ & $7.31\times10^{10}$ & $8.52\times10^{11}$ \\
    $g22795_0$ & $1.06\times10^{11}$ & $4.55\times10^{10}$ & $8.52\times10^{11}$ \\
	$g15784_0$ & $1.30\times10^{11}$ & $1.14\times10^{11}$ & $1.31\times10^{12}$ \\
    $g4720_0 $ & $1.42\times10^{11}$ & $5.51\times10^{10}$ & $1.02\times10^{12}$ \\
	$g4145_0 $ & $1.50\times10^{11}$ & $8.09\times10^{10}$ & $1.19\times10^{12}$ \\
	$g25271_0$ & $1.56\times10^{11}$ & $7.87\times10^{10}$ & $1.25\times10^{12}$ \\
	$g27491_0$ & $1.88\times10^{11}$ & $2.08\times10^{11}$ & $2.14\times10^{12}$ \\
	$g15807_0$ & $2.14\times10^{11}$ & $1.75\times10^{11}$ & $2.03\times10^{12}$ \\
	\hline
\end{tabular}